\documentclass[twocolumn,showpacs,preprintnumbers,amsmath,amssymb,footinbib]{revtex4-1}

\usepackage{graphicx}
\usepackage{dcolumn}
\usepackage{bm}
\usepackage{color}
\usepackage{soul}
\usepackage{pdfpages}

\begin{document}

\title{An acoustic black hole in a stationary hydrodynamic flow of microcavity polaritons}

\author{H.~S.~Nguyen$^1$}
\email{hai\_son.nguyen@lpn.cnrs.fr}
\altaffiliation[Present address: ]{Institut de Nanotechnologies de Lyon, Ecole Centrale de Lyon, CNRS (UMR 5270), 36 avenue Guy de Collongue, 69134 Ecully, France}
\author{D.~Gerace$^2$}
\author{I.~Carusotto$^3$}
\author{D.~Sanvitto$^4$}
\author{E.~Galopin$^1$}
\author{A.~Lema\^{i}tre$^1$}
\author{I.~Sagnes$^1$}
\author{J.~Bloch$^1$}
\author{A.~Amo$^1$}
\affiliation{$^1$Laboratoire de Photonique et de Nanostructures, LPN/CNRS, Route de Nozay, 91460 Marcoussis , France}
\affiliation{$^2$Dipartimento di Fisica, Universit\`a di Pavia, via Bassi 6, I-27100 Pavia, Italy}
\affiliation{$^3$INO-CNR BEC Center and Dipartimento di Fisica, Universit\`a di Trento, I-38123 Povo, Italy}
\affiliation{$^4$NNL, Istituto Nanoscienze - CNR, Via Arnesano, 73100 Lecce, Italy}

\date{\today}

\begin{abstract}
We report an experimental study of superfluid hydrodynamic effects in a one-dimensional polariton fluid flowing along a laterally patterned semiconductor microcavity and hitting a micron-sized engineered defect. 
At high excitation power, superfluid propagation effects are observed in the polariton dynamics, in particular, a sharp acoustic horizon is formed at the defect position, separating regions of sub- and super-sonic flow. Our experimental findings are quantitatively reproduced by theoretical calculations based on a generalized Gross-Pitaevskii equation. Promising perspectives to observe Hawking radiation via photon correlation measurements are illustrated.  
\end{abstract}

\maketitle

{\em Introduction -- } Back in 1974, S.~Hawking~\cite{Hawking74} predicted that the zero-point fluctuations of the quantum vacuum may be converted into correlated pairs of real particles at the event horizon of an astrophysical black hole. The resulting emission, called Hawking radiation, has a thermal character at a Hawking temperature $T_H$ proportional to the surface gravity of the black hole. Unfortunately, a direct observation of this radiation in an astrophysical context is made very difficult by the extremely low value of $T_H$, in the micro-K range for a solar mass black hole, well below the cosmic microwave background.

To overcome this severe limitation, a pioneering work by Unruh~\cite{Unruh81} introduced the idea of condensed-matter analogs of gravitational systems, and anticipated the occurrence of an analog Hawking emission of sound waves whenever a fluid shows an acoustic horizon separating regions of sub- and supersonic flow. Since this early prediction, an intense theoretical activity was devoted to the study of different condensed-matter and optical platforms where such analog black holes could be created~\cite{barcelo}. In the last few years, experimental realizations of horizons were reported in water channels~\cite{Weinfurtner}, atomic Bose-Einstein condensates \cite{Lahav}, nonlinear optical fibers~\cite{Philbin} and silica glass~\cite{Belgiorno} illuminated by strong laser pulses, and also in bulk nonlinear optical systems~\cite{BarAd}. However, the only experimental claims of stimulated and spontaneous Hawking emission reported so far respectively in~\cite{Weinfurtner} and~\cite{Belgiorno} are still being actively debated by the 
community~\cite{Rousseaux,Parentani14,Unruh_comment,Finazzi}.

In the last decade, quantum fluids of light~\cite{RMPIacopo} have emerged as a promising system to study quantum hydrodynamics effects, and in particular analog black holes and Hawking radiation. Among the different optical platforms, dressed photons in semiconductor microcavities in the strong coupling regime -- the so-called cavity polaritons \cite{Weisbuch} -- have led to the most impressive experimental achievements, such as the observation of superfluidity \cite{Amofluid} and of the hydrodynamic nucleation of quantized vortices \cite{Nardin,Sanvitto2011} and solitons \cite{Amosoliton,Sich} in a novel optical context. Building on these results, strategies to study acoustic Hawking emission from analog black holes in polariton fluids have been recently proposed~\cite{Marino,Fleurov,Dima,Dario,Larre}. 

In this Letter, we report the experimental realization of an acoustic black hole in the hydrodynamic flow of microcavity polaritons. A stationary one-dimensional (1D) {flowing} polariton fluid is generated by resonant cw excitation of a microcavity device laterally patterned into a photonic wire showing a localized defect potential. While at weak excitation power the polariton density profile shows a strong modulation due to scattering on the defect, at stronger excitation powers a superfluid behavior is apparent in the suppressed modulation upstream of the defect~\cite{Iacopofluid,Amofluid}; at the same time, as a consequence of the sudden density drop an acoustic horizon appears in the vicinity of the defect, after which the flow is supersonic. Detailed in-situ information on the flow is obtained from real- and momentum-space photoluminescence (PL) experiments, which show full quantitative agreement with theoretical predictions based on the generalized Gross-Pitaevskii equation of Ref.~\onlinecite{Iacopofluid}.

\begin{figure}[t]
\begin{center}
\includegraphics[width=8.5cm]{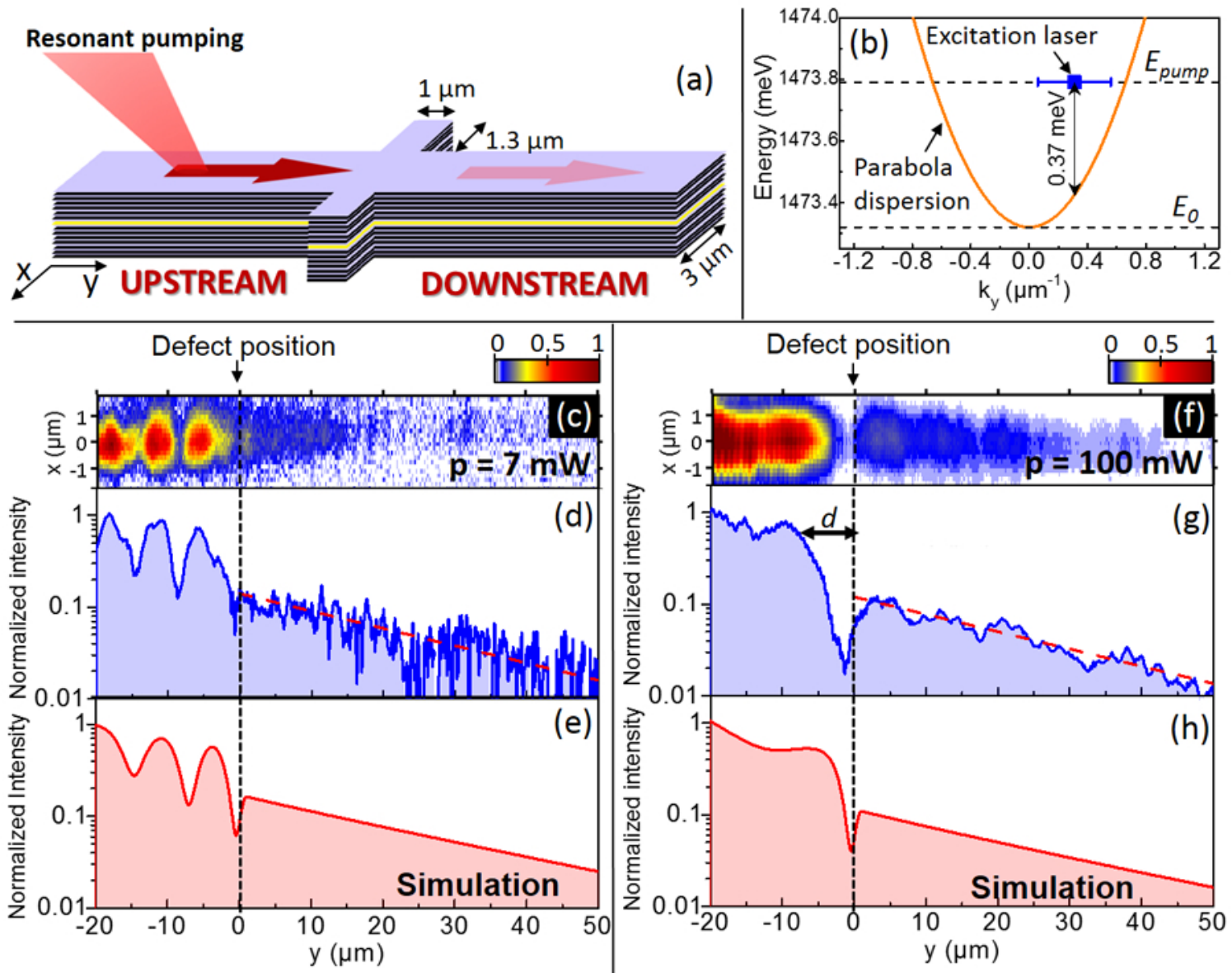}
\end{center} 
\caption{(Color online). (a) Sketch of the experimental configuration used to create a polariton fluid showing an acoustic horizon. (b) Laser excitation conditions in the wavevector-energy plane. (c,f) Linear scale image of the spatially resolved PL emission for excitation powers $p=7$ mW and $p=100$ mW, respectively. (d,g) Semilogarithmic plot of the PL emission integrated over the transverse direction. (e,h) Generalized Gross-Pitaevskii numerical simulation of the quantities in (d,g).}
\label{fig1}
\end{figure}

\textit{Experimental set-up -- }
Our sample, grown by molecular beam epitaxy, consists of a GaAs $\lambda$ microcavity sandwiched between top/bottom Bragg reflectors, with 26/30 pairs of Ga$_{0.9}$Al$_{0.1}$As/Ga$_{0.05}$Al$_{0.95}$As $\lambda/4$ layers. A single 8~nm-thick  In$_{0.95}$Ga$_{0.05}$As quantum well is inserted at the antinode of the confined electromagnetic field, i.e. at the center of the microcavity layer. At 10~K, the temperature of our experiments, the photon quality factor amounts to $Q\sim 33000$ (i.e. $\hbar\gamma\sim 0.05$~meV, $\gamma$ being the photon decay rate) and the Rabi splitting is $\hbar\Omega_R\sim3.5$~meV. Electron beam lithography and inductively coupled plasma dry etching were used to laterally pattern the microcavity, such that photonic wires of $400$ $\mu$m-length and $3$ $\mu$m-width were realized. In the considered spatial region, the detuning of the photonic wire mode from the exciton energy is $\delta=E_{\mathrm{cav}}-E_{\mathrm{ex}}\approx -3$~meV, corresponding to a lower polariton effective mass $3\times 10^{-5}$ times smaller than the free electron mass. 
An engineered defect is created at the center of the wire by enlarging its width to $5.6$ $\mu$m over a length of $1$ $\mu$m, as schematically shown in Fig.~\ref{fig1}(a): the reduced lateral confinement results in a localized attractive potential for polaritons with a depth of $-0.85$~meV. 

In our experiments, a cw single-mode Ti:Sapphire excitation laser is focused onto a $16$ $\mu$m-wide spot by means of a microscope objective with numerical aperture NA$=0.28$. The laser spot is centered about 30~$\mu$m away from the defect, and the incidence angle corresponds to a laser wave vector around $k_{\mathrm{p}}=0.3$ $\mu$m$^{-1}$. The laser energy $E_{\mathrm{pump}}$ is blue-shifted by 0.37~meV with respect to the (approximately) parabolic lower polariton dispersion, as shown in Fig.~\ref{fig1}(b): As a result, polaritons ballistically propagate out of the excitation spot towards the defect~\cite{RMPIacopo}, as schematically illustrated in Fig.~\ref{fig1}(a). Polariton emission, collected in transmission geometry through the back of the sample with a second microscope objective (NA$=0.55$), is imaged on a charge-coupled device (CCD) camera coupled to a spectrometer; both real and momentum space images are acquired.

\begin{figure}[t]
\begin{center}
\includegraphics[width=8.0cm]{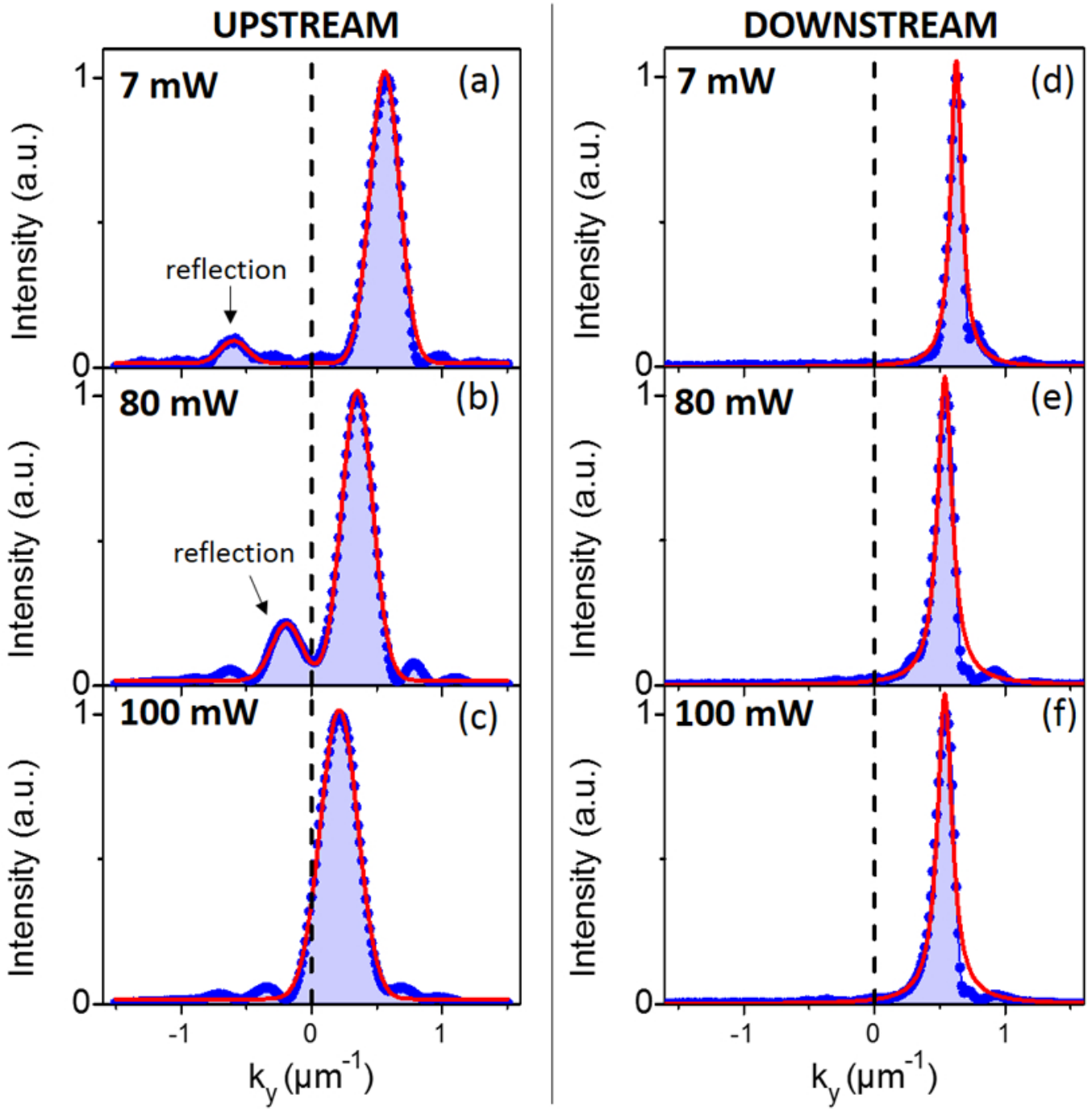}
\end{center}
\caption{(Color online). Wavevector-resolved PL intensity for different excitation powers. (a-c) A spatial filtering from $y=-25\,\mu m$ to $y=-5\,\mu m$ is used to isolate the signal from the upstream region. (d-f) Same plots with a spatial filtering from $y=5$ $\mu$m to $y=50$ $\mu$m to isolate the signal from the downstream region. Blue dots are the experimental points, red lines are Gaussian fits.}
\label{fig2}
\end{figure}

{\em Low-power experiments --} The polariton flow across the defect is first characterized at low excitation power ($p=7$~mW). At such low power, polariton-polariton interactions are negligible and the Bogoliubov dispersion of the elementary excitations in the fluid~\cite{Iacopofluid,RMPIacopo} reduces to the single particle parabolic dispersion: back-scattering of polaritons from the defect is thus possible, and interference between the incoming and reflected polaritons is visible in the spatially resolved PL emission of the polariton fluid as a strong modulation of the density pattern in the region upstream of the defect [Fig.~\ref{fig1}(c)]. The corresponding momentum space distribution is shown in Fig.~\ref{fig2}(a) where the light emission from the upstream region is isolated using a spatial filter~\cite{Supp}: the peak at $k_y>0$ corresponds to the incoming fluid, while the one at $k_y<0$ is due to reflected polaritons. An analog measurement on the other side of the defect shows that the polaritons recover a ballistic flow with a single momentum space peak at $k_y>0$ [Fig.~\ref{fig2}(d)], while the spatial density exponentially decreases because of radiative losses [Fig.~\ref{fig1}(d)]: the radiative lifetime estimated from the measured polariton momentum and spatial decay rate amounts to 14~ps (corresponding to $\hbar \gamma = 47\,\mu eV$), in very good agreement with the nominal quality factor of our sample. 



{\em High-power experiments -- } As reviewed in~\cite{barcelo}, the very concept of analog models is based on the possibility to describe the wavy propagation of excitations on the fluid in terms of a curved space-time metric. In particular, this requires that excitations have a linear, sonic-like dispersion on both sides of the horizon, the speed of sound then playing the role of the speed of light. While the low-density excitation spectrum of the polariton fluid has a single-particle parabolic shape, a sonic dispersion is recovered at higher densities, with a speed of sound given by $c=\sqrt{\hbar g n/m}$ in terms of the polariton-polariton interaction energy, $\hbar g$, and the polariton density, $n$~\cite{Iacopofluid,Kohnle}. In a flowing fluid at velocity $v$, density excitations are dragged by the flow and propagate at $v\pm c$. An acoustic horizon is defined as the point where a spatially dependent flow changes from a subsonic ($v<c$) character in the upstream region to a supersonic ($v>c$) character in the downstream one: in such configurations, sonic excitations are unable to propagate back from the supersonic region to the horizon, analogously to what happens to light trying to escape from astrophysical black holes.

In the experiment, acoustic black-hole horizons are created by increasing the laser power so to inject a higher polariton density while keeping the same excitation geometry and laser energy. Experimental results for a $100$~mW excitation power are shown in Fig.~\ref{fig1}(f): in contrast to the low density case, the reflection on the defect is now totally suppressed, as clearly evidenced by the absence of any interference pattern in the upstream region. Correspondingly, the $k_y<0$ peak shown in Fig.~\ref{fig2}(c) disappears from the spatially-selected momentum space data for the upstream region. As originally studied in~\cite{Iacopofluid,Amofluid}, these features are a clear signature of the superfluid nature of the polariton fluid in the upstream and of the sub-sonic character of the flow. 

As one can see in Fig.~\ref{fig1}(g), the density of the polariton flow drops by a factor of $\sim 7$ over a distance $d\approx 8$~$\mu$m across the defect region, with an even deeper minimum a few $\mu$m in front of the defect. This complicate density profile is quantitatively reproduced by numerical simulations of a generalized Gross-Pitaevskii equation for the lower polariton field including pumping and loss terms~\cite{Iacopofluid,RMPIacopo} as shown in Fig.~\ref{fig1}(h). The parameters used in the simulations (potential profile, shape and energy of the pump, losses) are directly taken from the nominal experimental values; more details on these calculations are given in the Supplemental Materials~\cite{Supp}. The density drop across the defect results in a corresponding decrease of the speed of sound and simultaneous increase of the flow speed, as evidenced by comparing the positions of the momentum-space peaks on the two sides of the defect shown in Figs.~\ref{fig2}(c,f).

\begin{figure}[t]
\begin{center}
\includegraphics[width=8.5cm]{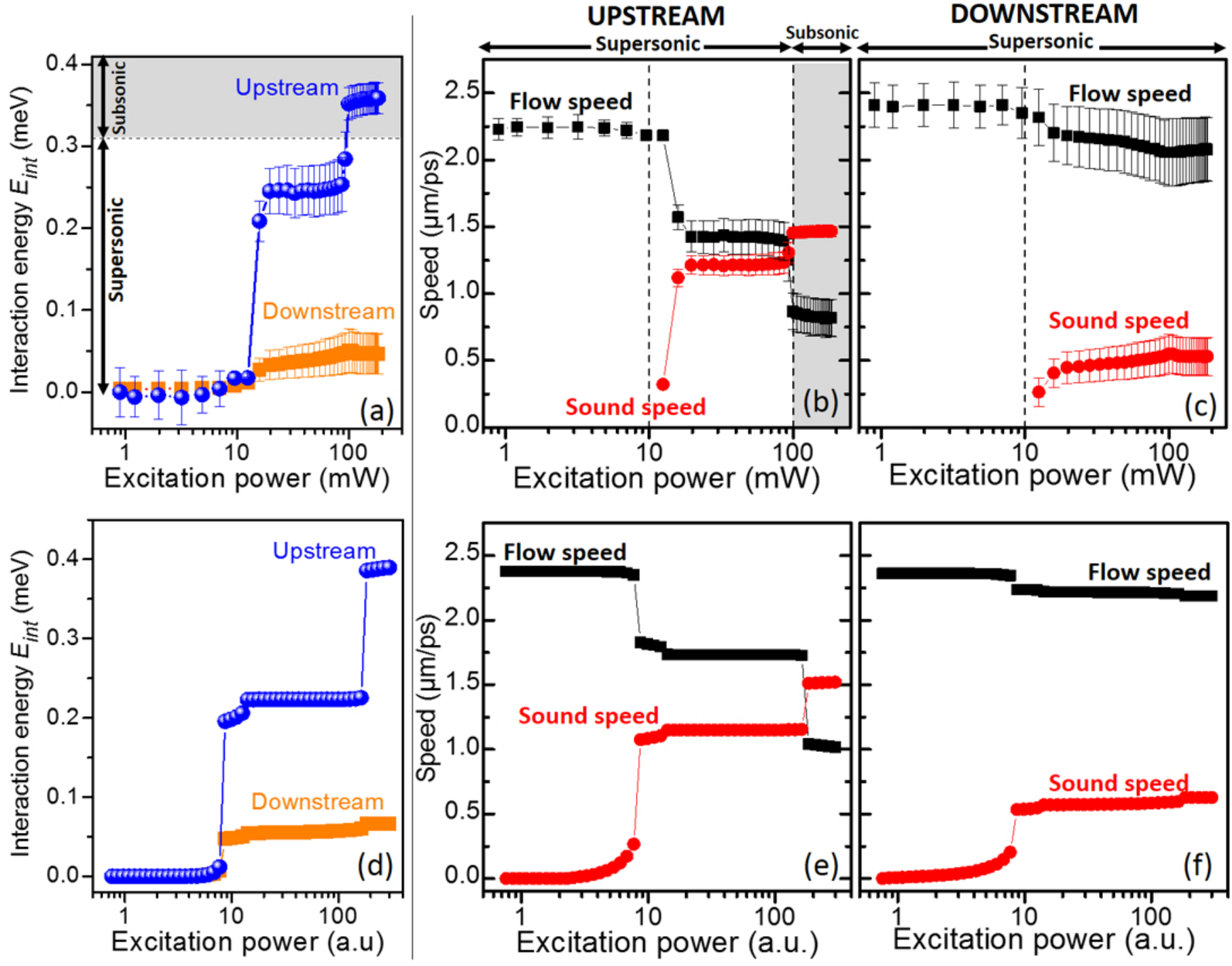} 
\end{center}
\caption{(Color online). (a) Interaction energy in the upstream (orange squares) and downstream (blue circles) regions as a function of excitation power. (b,c) Flow speed and speed of sound in the (b) upstream and (c) downstream regions as a function of excitation power. (d,e,f) Results of a generalized Gross-Pitaevskii numerical simulation of the same quantities. The procedure discussed in the text is used to extract the interaction energy, the flow speed, and the speed of sound for both experimental data and theoretical calculations.}
\label{fig3}
\end{figure}

{\em Experimental evidence of the acoustic horizon -- } A quantitative insight on the subsonic vs. supersonic character of the flow in the upstream/downstream regions as a function of excitation power can be obtained from the momentum space pictures of Fig.~\ref{fig2} as follows. In our non-equilibrium stationary state, the polariton oscillation frequency is locked to the excitation laser frequency~\cite{Iacopofluid}. Away from the pump spot, in regions of slowly varying flow where quantum pressure is negligible, the Gross-Pitaevskii equation directly leads to a generalized Bernoulli equation for driven-dissipative polariton fluids in the form~\cite{RMPIacopo}
\begin{equation}
E_{\mathrm{pump}}=E_0+ \hbar^2 k_{\mathrm{fluid}}^2 /2m +
E_{\mathrm{int}} \,
\label{eq1}
\end{equation}
where $E_0$ is the single-particle polariton energy at $k_y=0$ [see Fig.~\ref{fig1}(b)], $\hbar k_{\mathrm{fluid}}$ is the local momentum of the polariton fluid, and $E_{\mathrm{int}}=\hbar g n$ is the interaction energy. {The physical meaning of this equation is visible in the momentum space data displayed in Fig.~\ref{fig2}(a-c): for the upstream region, the increase of the density $n$ with excitation power corresponds to a decreased kinetic energy, indeed, as evidenced by the peak wavevector shifting from $k_{\mathrm{fluid}}=0.56$ $\mu$m$^{-1}$ (non-interacting fluid) to $k_{\mathrm{fluid}}=0.21$ $\mu$m$^{-1}$ (high power). Using the Eq.~(\ref{eq1}), it is then straightforward to extract from the peak wavevector, $k_{\mathrm{fluid}}$, the dependence of the interaction energy, $E_{\mathrm{int}}$, on the laser excitation power. The speed of sound is directly obtained from the interaction energy ($c=\sqrt{E_{\mathrm{int}}/m}$). The same analysis can be done for the downstream region [Fig.~\ref{fig2}(d-f)]. The results are shown in Fig.~\ref{fig3}. We distinguish three regimes as follows. }

(1) \textit{Linear regime} (for an excitation power $p<10$~mW). In this case, $E_{\mathrm{int}}\simeq 0$ in both the upstream and the downstream regions. Correspondingly, the flow speed ($v=\hbar k_{\rm fluid}/m$) is quite large and almost constant everywhere as it is fixed by the pump frequency~\cite{RMPIacopo}.

(2) \textit{Intermediate regime with supersonic-supersonic interface} (for 10~mW$<p<$100~mW). In this case, the polariton density is large enough to have a well-defined speed of sound in both upstream and downstream regions. After a fast increase at low $p$, the interaction energy saturates above $p=20$~mW. This can be explained in terms of an optical limiter behavior in the pump spot region, which sets in as soon as the interaction energy here exceeds the excitation laser detuning (0.37~meV). It is interesting to note that this value of the interaction energy is large enough ($E_{\mathrm{int}} > 2(E_{\mathrm{pump}} -E_0)/3\approx 0.31$~meV~\cite{Supp}) to give a sub-sonic superfluid flow under the laser spot. However, outside the excitation spot but still in the upstream region, the decrease in density caused by the radiative decay quickly turns the fluid back into the supersonic regime with $E_{\mathrm{int}} < 0.31$~meV, as shown in Fig.~\ref{fig3}(a-b).

(3) \textit{Subsonic-supersonic interface} (for $p>100$~mW). In this case, the polariton density in the upstream region between the pump spot and the defect is large enough $E_{\rm int}\approx 0.35\pm0.02$~meV to have a sub-sonic flow (superfluid) up to the defect position, in agreement with the un-modulated spatial shape already seen in Fig.~\ref{fig1}(g). On the other hand, as a consequence of the density drop at the defect position, the polariton density in the downstream region remains low enough for the flow to be supersonic, as shown in Fig.~\ref{fig3}(c). This transition between subsonic and supersonic regimes shows that for an excitation power $p>100$~mW, an acoustic black-hole horizon is indeed present in the polariton fluid in the vicinity of the engineered defect. From the quite sharp spatial profiles of the flow velocity and of the speed of sound, we can anticipate a quite large analog Hawking temperature $T_H$, in the Kelvin range~\cite{Supp}.

\begin{figure}[t]
\begin{center}
\includegraphics[width=0.4\textwidth]{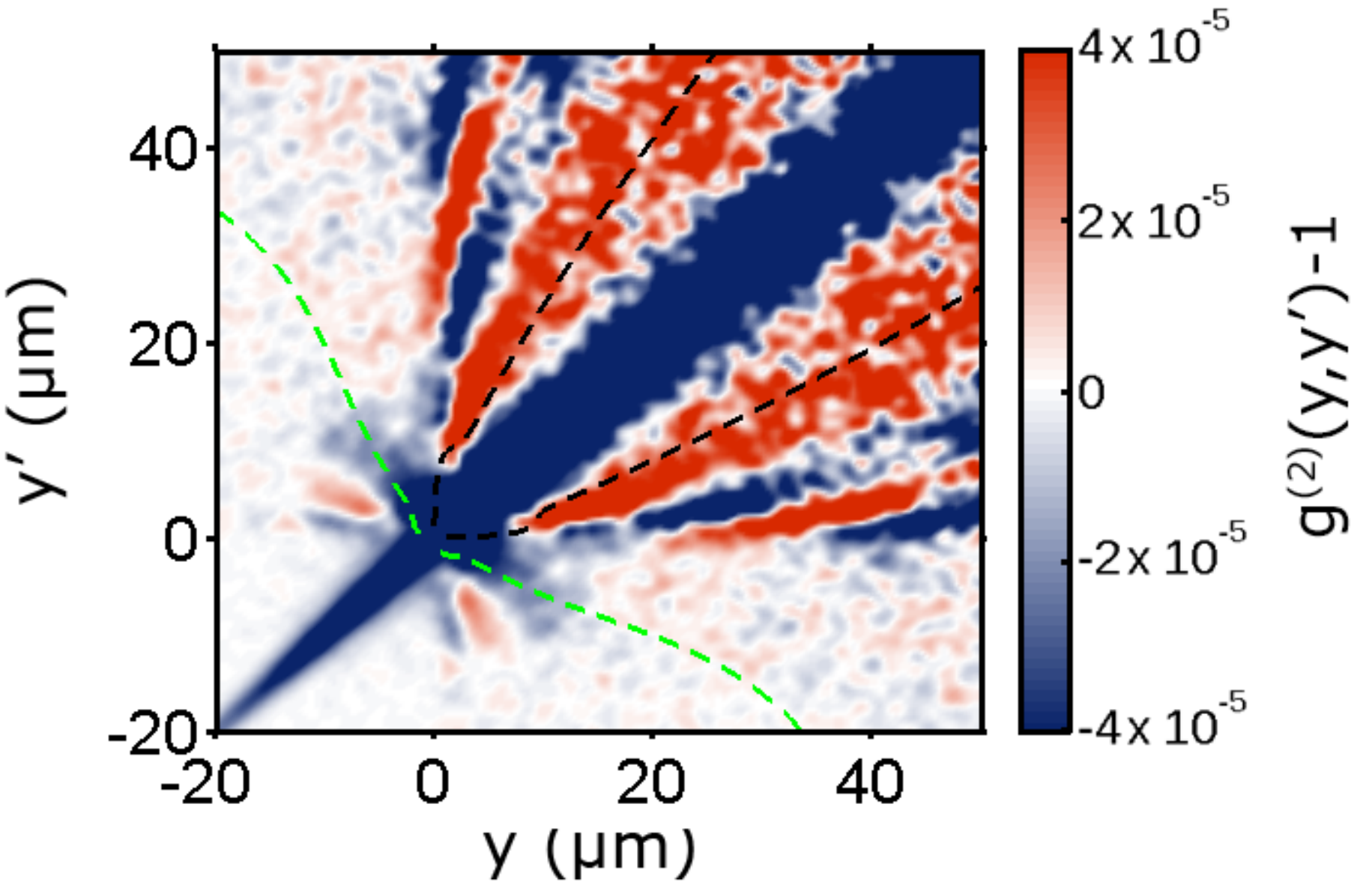} 
\end{center}
\caption{(Color online). Normalized spatial correlation function of photon density fluctuations, $g^{(2)}(y,y')-1$, as predicted by an out-of-equilibrium Wigner Monte Carlo simulation~\cite{Dario} using the experimental parameters. A gaussian real-space smoothening with $\ell_{\rm av}=1\,\mu$m has been used.}
\label{fig4}
\end{figure}

\textit{How to detect analog Hawking emission --} 
Finally, we address the actual possibility of unequivocally detecting the spontaneous Hawking radiation in our polariton acoustic black hole. As originally predicted in~\cite{BalbinotPRA}, the correlations between the Hawking partners can be detected in the spatially resolved, second-order correlation function of the density, i.e. 
\begin{equation}
g^{(2)}(y,y')\equiv \frac{\langle : n(y) \, n(y') : \rangle }{\langle n(y) \rangle \langle n(y')\rangle}  \, .
\end{equation} 
Such a signature of the analog Hawking radiation was shown to extend to the driven-dissipative case of a polariton fluid: as fluctuations of the in-cavity polariton density are directly observable as intensity fluctuations of the emitted light, the Hawking signal can be observed in the equal-time correlations of the spatially-resolved intensity noise of the emission~\cite{Dario}.

To estimate the feasibility of such a measurement, we have performed a Wigner Monte Carlo simulation of the polariton field dynamics along the same lines of Ref.~\onlinecite{Dario} (see the Supplemental Materials~\cite{Supp} for details) with the nominal parameters of this experiment. In particular, we have assumed the value  $\hbar g=0.3$ $\mu$eV$\cdot\mu$m for the 1D polariton-polariton interaction as extracted from another experience on the same sample \cite{bistability}. The result of an average over $2\times 10^5$ Montecarlo realizations is shown in Fig.~\ref{fig4}. Correlation and anti-correlation signals are clearly visible between points belonging to different regions along the wire axis. 
{The dashed lines indicate the points at which one would expect correlations to be strongest: compared to ~\cite{IacopoBH}, their curved shape stems from the spatial dependence of the flow and sound velocities due to the finite polariton lifetime (see the Supplemental Materials~\cite{Supp} for more details). For the same reason, all correlation signals only extend for a finite distance from the horizon~\cite{Dario}: if needed, this distance can be extended by working with an improved sample, e.g. with a larger quality factor cavity providing longer polariton lifetime. Finally, the additional fringes are a consequence of the curvature of the Bogoliubov dispersion, and of the complex density and velocity profiles in proximity to the horizon region~\cite{Zapata}. Even though the expected correlation signal is intrinsically quite low, an integration of the emitted radiation over macroscopically long times, i.e. many orders of magnitude longer than the polariton lifetime} and {in structures with reduced size and improved designs~\cite{Zapata}, {opens promising perspectives in view of its experimental detection.}
                                                                                                  
\textit{Conclusions -- } In this Letter we have reported an experimental study of superfluid hydrodynamics of polaritons along a one-dimensional microstructure including an engineered defect. In particular, we have demonstrated the formation of an acoustic black hole horizon. Theoretical Wigner Monte Carlo calculations show that this configuration is amenable to the detection of a Hawking radiation signal in the spatially-resolved correlation function of the intensity noise of photoluminescence, a quantity directly accessible to quantum optical techniques~\cite{BalbinotPRA,IacopoBH,Dario}. 
In contrast to most other analog models where the high-energy part of the dispersion affects the properties of the zero-point emission~\cite{Finazzi,Parentani14}, excitations on top of the polariton fluid satisfy the hypothesis of the gravitational analogy and are well described by a quantum field theory on a curved space-time~\cite{birrell,barcelo}. Moreover, with respect to atomic systems, both the high value of $T_H$ in the few-K range and the possibility of very long integration times offered by the stationary nature of the polariton flow are very promising in view of an experimental observation of the Hawking emission signal. Standard optical and nanotechnology tools can be used to engineer the spatial shape of the horizon~\cite{RTD}, thus reinforcing the Hawking signal~\cite{Zapata}. Further useful experimental knobs are offered by the polariton spin degrees of freedom~\cite{Dima,Larre}. Given this remarkable flexibility in the all-optical generation and manipulation of polariton fluids, our results suggest them as a unique tool to study a variety of analogue gravity effects in novel and tunable experimental conditions.

The authors acknowledge R. Balbinot, D.~Ballarini, A.~Bramati, P.-\'E. Larr\'e, N.~Pavloff, R.~Parentani, F.~Sols, and I.~Zapata for fruitful discussions. This work was supported by the ANR Project ``Quandyde" (No. ANR-11-BS10-001), by the French RENATECH network, the LABEX Nanosaclay, the ERC grants Honeypol and QGBE, and the Autonomous Province of Trento, partly under the call {\em ``Grandi Progetti 2012"}, project {``On silicon chip quantum optics for quantum computing and secure communications - SiQuro"}.


\begin{thebibliography} {99}

\bibitem{Hawking74} S. W. Hawking, Nature \textbf{248}, 30 (1974).
\bibitem{Unruh81} W. G. Unruh, Phys. Rev. Lett. \textbf{46}, 1351 (1981).
\bibitem{barcelo}  C. Barcel\'o, S. Liberati, and M. Visser,  Living Rev. Relativity {\bf 8}, 12 (2005). 
\bibitem{Weinfurtner} S. Weinfurtner, E. W. Tedford, M. C. J. Penrice, W. G. Unruh, and G. A. Lawrence, Phys. Rev. Lett. \textbf{106}, 021302 (2011).
\bibitem{Lahav} O. Lahav, A. Itah,  A. Blumkin, C. Gordon, S. Rinott, A. Zayats, and J. Steinhauer, Phys. Rev. Lett. \textbf{105}, 240401 (2010).
\bibitem{Philbin} T. G. Philbin, C. Kuklewicz, S. Robertson, S. Hill, F. K\"onig, and U. Leonhardt, Science \textbf{319}, 5868 (2008).
\bibitem{Belgiorno} F. Belgiorno, S. L. Cacciatori, M. Clerici, V. Gorini, G. Ortenzi, L. Rizzi, E. Rubino, V. G. Sala, and D. Faccio, Phys. Rev. Lett. \textbf{105}, 203901 (2010).
\bibitem{BarAd} M. Elazar, V. Fleurov, and S. Bar-Ad, Phys. Rev. A \textbf{86}, 063821 (2012).
\bibitem{Unruh_comment} R. Sch\"{u}tzhold and W. G. Unruh, Phys. Rev. Lett. \textbf{107}, 149401 (2011).
\bibitem{Finazzi}  S. Finazzi and I. Carusotto, Phys. Rev. A \textbf{89}, 053807 (2014).
\bibitem{Parentani14} F. Michel and R. Parentani, Phys. Rev. D {\bf 90}, 044033 (2014).
\bibitem{Rousseaux} L.-P. Euv\'e, F. Michel, R. Parentani, and G. Rousseaux, preprint arXiv:1409.3830.

\bibitem{RMPIacopo}  I. Carusotto and C. Ciuti, Rev. Mod. Phys. \textbf{85}, 299 (2013).
\bibitem{Weisbuch}  C. Weisbuch, M. Nishioka, A. Ishikawa, and Y. Arakawa, Phys. Rev. Lett. \textbf{69}, 3314 (1992).
\bibitem{Amofluid}  A. Amo, J. Lefr\`ere, S. Pigeon, C. Adrados, C. Ciuti, I. Carusotto, R. Houdr\'e, E. Giacobino, and A. Bramati, Nat. Physics \textbf{5}, 805 (2009).
\bibitem{Nardin} G. Nardin, G. Grosso, Y. Leger, B. Pietka, F. Morier-Genoud, and B. Deveaud-Pledran, Nat. Physics \textbf{7}, 635 (2011).
\bibitem{Sanvitto2011} 
D. Sanvitto, S. Pigeon, A. Amo, D. Ballarini, M. D. Giorgi, I. Carusotto, R. Hivet, F. Pisanello, V. G. Sala, P. S. S. Guimar\~{a}es, R. Houdr\'e, E. Giacobino, C. Ciuti, A. Bramati, and G. Gigli, Nat. Photonics \textbf{5}, 610 (2011).
\bibitem{Amosoliton} 
A. Amo, S. Pigeon, D. Sanvitto, V. G. Sala, R. Hivet, I. Carusotto, F. Pisanello, G. Lem\'{e}nager, R. Houdr\'e, E. Giacobino, C. Ciuti, and A. Bramati, Science \textbf{332}, 1167 (2011).
\bibitem{Sich} 
M. Sich, D. N. Krizhanovskii, M. S. Skolnick, A. V. Gorbach, R. Hartley, D. V. Skryabin, E. A. Cerda-M\'{e}ndez, K. Biermann, R. Hey, and P. V. Santos, Nat. Photonics \textbf{6}, 50 (2012).
\bibitem{Marino} F. Marino, Phys. Rev. A {\bf 78}, 063804 (2008). 
\bibitem{Fleurov} I. Fouxon, O. V. Farberovich, S. Bar-Ad, and V. Fleurov, Europhys. Lett. {\bf 92}, 14002 (2010).
\bibitem{Dima}  D. D. Solnyshkov, H. Flayac, and G. Malpuech, Phys. Rev. B \textbf{84}, 233405 (2011).
\bibitem{Dario} D. Gerace and I. Carusotto, Phys. Rev. B  \textbf{86}, 144505 (2012).
\bibitem{Larre} P. \'E. Larr\'e, N. Pavloff, and A. M. Kamchatnov, Phys. Rev. B \textbf{88}, 224503 (2013).
\bibitem{Iacopofluid}  I. Carusotto and C. Ciuti, Phys. Rev. Lett. \textbf{93}, 166401 (2004).
\bibitem{Supp} Supplemental Material.

\bibitem{Kohnle} V. Kohnle, Y. L\'eger, M. Wouters, M. Richard, M. T. Portella-Oberli, and B. Deveaud-Pl\'edran, Phys. Rev. Lett. {\bf 106}, 255302 (2011).
 \bibitem{bistability} H. S. Nguyen \textit{et al.}, in preparation. 
\bibitem{IacopoBH}  I. Carusotto, S. Fagnocchi, A. Recati, R. Balbinot, and A. Fabbri, New J. Phys. \textbf{10}, 103001 (2008).
\bibitem{BalbinotPRA}  R. Balbinot, A. Fabbri, S. Fagnocchi, A. Recati, and I. Carusotto, Phys. Rev. A  {\bf 78}, 021603(R) (2008).
\bibitem{Zapata}  I. Zapata, M. Albert, R. Parentani, and F. Sols, New J.Phys. \textbf{13}, 063048 (2011).
\bibitem{birrell} N. D. Birrell and P. C. W. Davies, {\em Quantum fields in curved space} (Cambridge University Press, Cambridge, UK, 1982).
\bibitem{RTD}  
H. S. Nguyen, D. Vishnevsky, C. Sturm, D. Tanese, D. Solnyshkov, E. Galopin, A. Lema\^{i}tre, I. Sagnes, A. Amo, G. Malpuech, and J. Bloch, 
Phys. Rev. Lett. \textbf{110}, 236601 (2013).


\end{thebibliography}
\end{document}


\title{Supplemental Information: An acoustic black hole in a stationary hydrodynamic flow of microcavity polaritons}

\author{H.~S.~Nguyen$^1$}
\email{hai\_son.nguyen@lpn.cnrs.fr}
\altaffiliation[Present address: ]{Institut de Nanotechnologies de Lyon, Ecole Centrale de Lyon, CNRS (UMR 5270), 36 avenue Guy de Collongue, 69134 Ecully, France}
\author{D.~Gerace$^2$}
\author{I.~Carusotto$^3$}
\author{D.~Sanvitto$^4$}
\author{E.~Galopin$^1$}
\author{A.~Lema\^{i}tre$^1$}
\author{I.~Sagnes$^1$}
\author{J.~Bloch$^1$}
\author{A.~Amo$^1$}
\affiliation{$^1$Laboratoire de Photonique et de Nanostructures, LPN/CNRS, Route de Nozay, 91460 Marcoussis , France}
\affiliation{$^2$Dipartimento di Fisica, Universit\`a di Pavia, via Bassi 6, I-27100 Pavia, Italy}
\affiliation{$^3$INO-CNR BEC Center and Dipartimento di Fisica, Universit\`a di Trento, I-38123 Povo, Italy}
\affiliation{$^4$NNL, Istituto Nanoscienze - CNR, Via Arnesano, 73100 Lecce, Italy}

\date{\today}

\maketitle

\renewcommand{\theequation}{S\arabic{equation}}
\renewcommand{\thefigure}{S\arabic{figure}}

\section{Sample characterization}

\begin{figure}
\begin{center}
\includegraphics[width=8.5cm]{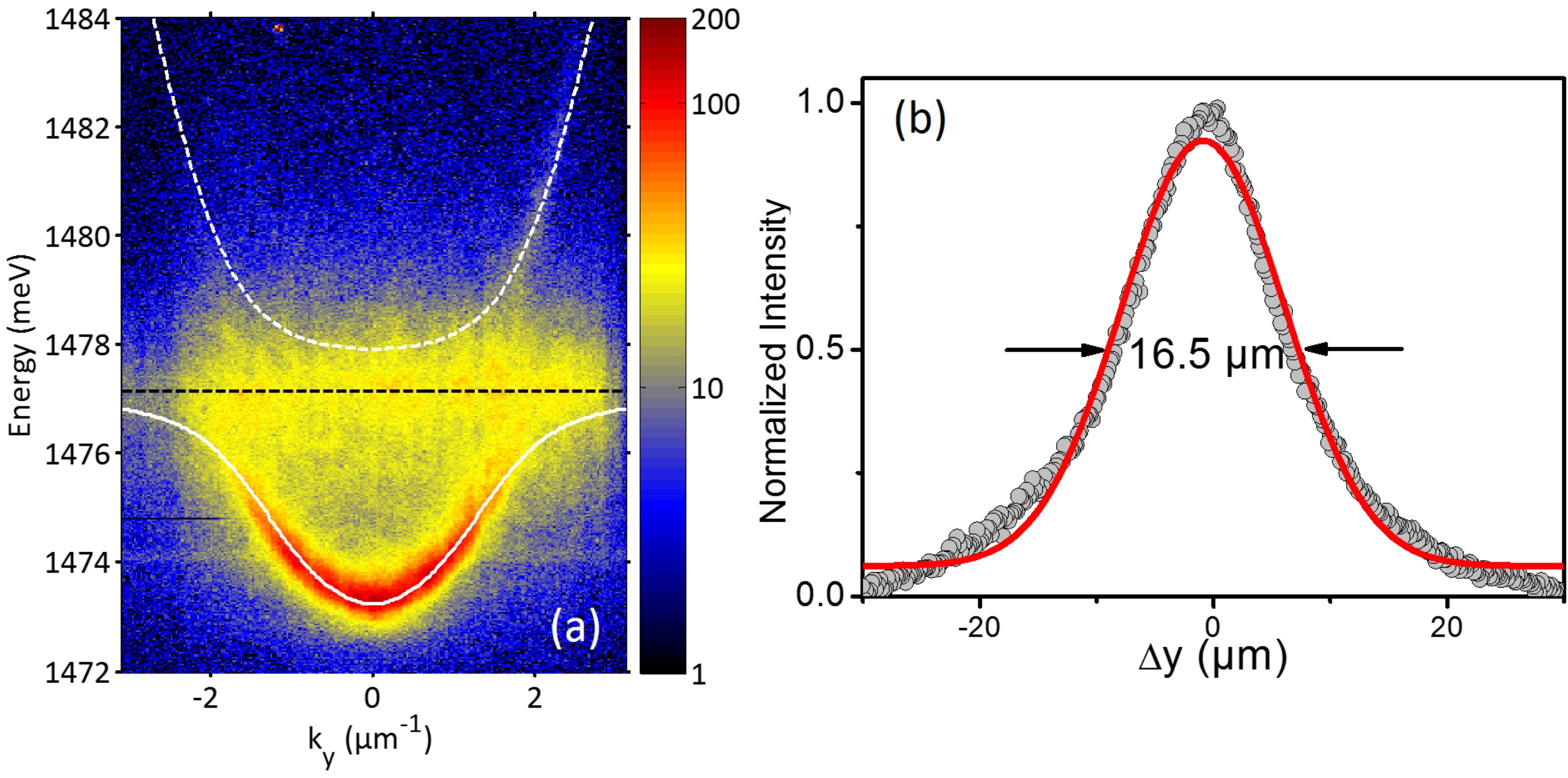}
\end{center}
\caption{\label{S1}{(a) Photoluminescence of the polariton wire under non-resonant excitation, showing the linear polariton dispersion. Dotted-black line: bare exciton energy. 
Dotted-white line (solid-white line): fit of the lower (upper) polariton branch from a two-coupled oscillators model. 
(b) Measured spatial profile of the integrated photoluminescence emission under non-resonant excitation. Red line: gaussian fit with FWHM = 16.5~$\mu$m. }}
\end{figure}

The polariton dispersion in the upstream and downstream regions is measured via non-resonant photoluminescence of the flat parts of the microstructure [Fig.~\ref{S1}]. The excitation energy is $72$~meV blue-detuned with respect to the exciton energy, $E_{\mathrm{ex}}=1477.7$~meV. Figure~\ref{S1}(a) shows the photoluminescence from both the polariton modes and the uncoupled excitons emitting through the wire-edge. From the photoluminesce spectrum, we extract a detuning  $\delta=E_{\mathrm{cav}}-E_{\mathrm{ex}}\approx -3$~meV, and a polariton effective mass $m\sim 3 \times10^{-5}m_\mathrm{el}$ (for a parabolic approximation of the lower polariton branch around $k=0$), where $m_\mathrm{el}$ represents the free electron mass.   

To characterize the shape of the excitation spot, we measure the spatial profile of the total emission along the wire axis, at very low excitation power, as depicted in Fig.~\ref{S1}(b). We use a gaussian spot with a full width at half maximum of~$\sim 16.5$ $\mu$m.

\section{Wave-vector measurements with spatial filtering}

\begin{figure}
\begin{center}
\includegraphics[width=8.5cm]{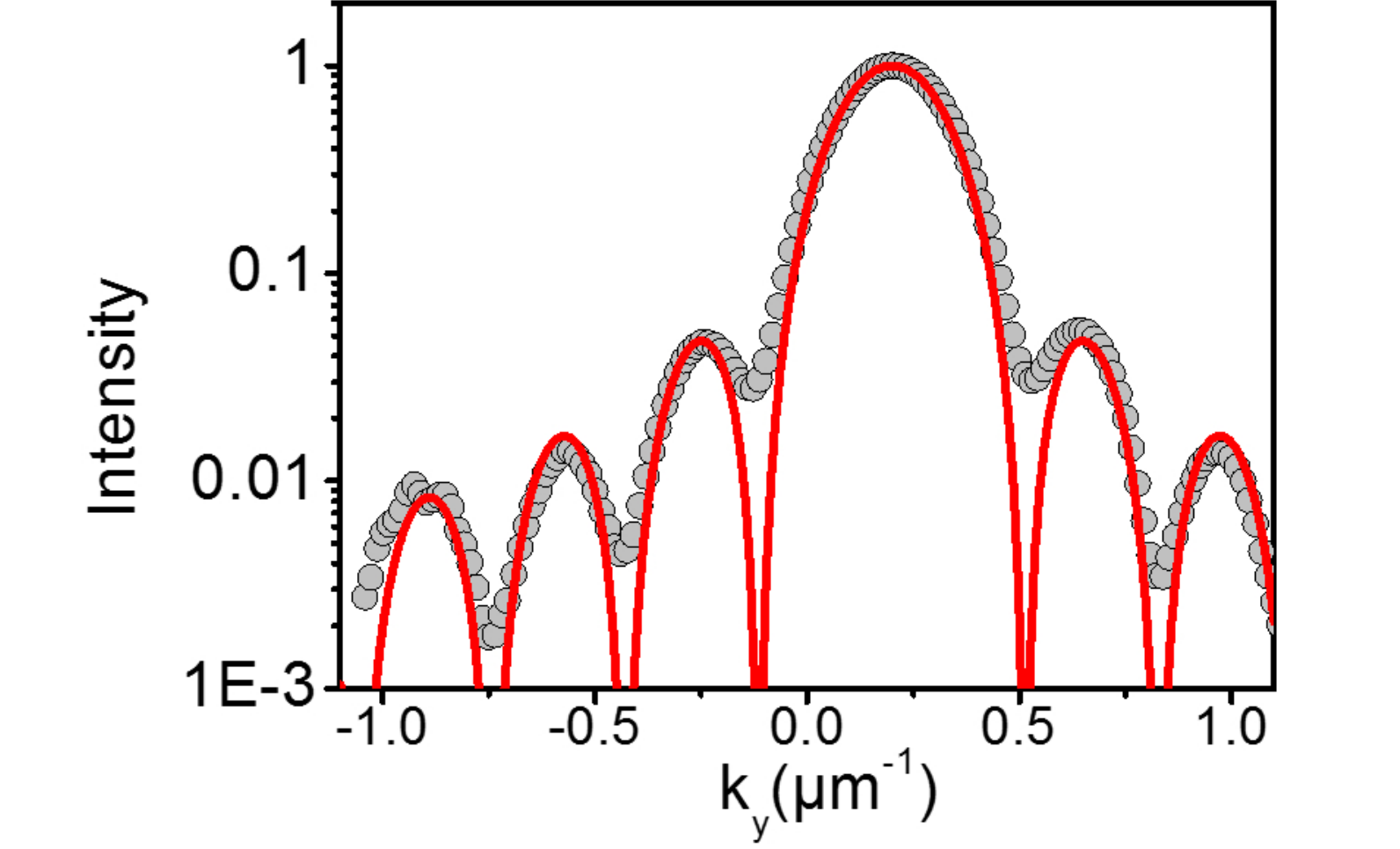}
\end{center}
\caption{\label{S2}{Measured momentum distribution of the superfluid in the upstream region at pump power $p=100$ mW [see also Fig.~2(c) of the main text], presented in semilogarithmic-scale. Red line: fit to a Fraunhofer diffraction response through a $20$ $\mu$m-wide slit. 
The fitting parameters are $w = 20.3$ $\mu$m and $k_{\mathrm{fluid}} = 0.2$ $\mu$m.}}
\end{figure}

In order to measure the selected emission in momentum space from the upstream or downstream regions, we spatially filter the emission from the desired region with the use of a slit located onto an intermediate image of the wire. The use of the slit induces a diffraction effect in the far-field images, particularly when we use a filter of $20$ $\mu$m in width, which is the case in the upstream measurements. This explains the broadening of the peaks in Figs.~2(a-c) of the main text compared to those in Figs.~2(d-f), for which the filter was $>45$ $\mu$m wide. Diffraction features are clearly evidenced in Fig.~\ref{S2} which represents the data of Fig.~2(c) in logarithmic scale. We observe not only a central peak but also side peaks that are perfectly fitted by a Fraunhaufer diffraction formula: $I=\sin^2\left[{w}(k_{\mathrm{fluid}}-k_y) /2 \right]/\left[{w}(k_{\mathrm{fluid}}-k_y)/{2}\right]^2$ with $w=20.3$ $\mu$m and $k_{\mathrm{fluid}}=0.2$ $\mu $m$^{-1}$. This shows that the side peaks in Fig.~2(c) do not originate from the backscattering of polaritons but are due to the diffraction of the central peak by the spatial selection slit.

\section{Interaction energy in the superfluid regime}

The superfluid regime is achieved if the polariton fluid is subsonic ($v<c$). This condition is equivalent to:
\begin{equation}
E_{\mathrm{int}} > 2E_{\mathrm{kinetic}} \, ,
\label{eqS2}
\end{equation}
where $E_{\mathrm{int}}=mc^2$ is the interaction energy and $E_{\mathrm{kinetic}}={m v^2}/{2}$ is the kinetic energy of the polariton fluid in the single-particle picture. Moreover, as discussed in the main text, the total polariton energy is fixed by the excitation laser frequency~\cite{RMPIacopo}, which implies that
\begin{equation}
E_{\mathrm{pump}}=E_0+ E_{\mathrm{kinetic}} + E_{\mathrm{int}} \, ,
\label{eqS3}
\end{equation} 
where $E_0$ is the single-particle polariton energy at $k_y=0$ [see Fig.~1(b) of the main text]. Therefore, in the superfluid regime the interaction energy satisfies:
\begin{equation}
E_{\mathrm{int}}>\frac{2}{3}\left(E_{\mathrm{pump}}-E_0\right) \, 
\label{eqS4}
\end{equation}

\begin{figure}[t]
\begin{center}
\includegraphics[width=\columnwidth]{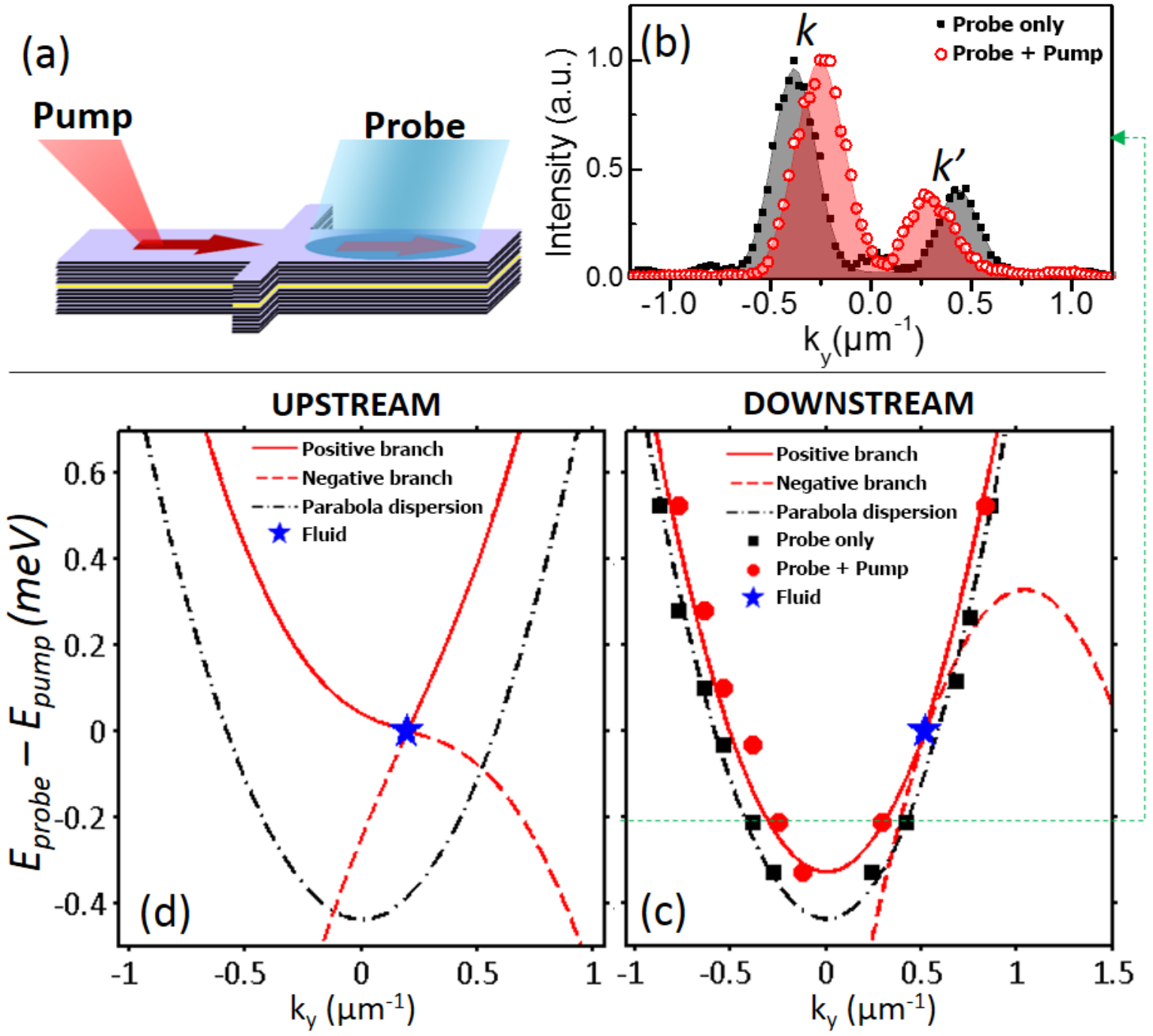}
\end{center}
\caption{\label{S3}{(a) Sketch of the pump-probe experiment to measure the dispersion of excitations in the downstream region of the polariton fluid. (b) Momentum distribution of the photoluminescence at the probe frequency with (empty red circles) and without (black squares) the pump. The probe energy is fixed as $E_{\mathrm{probe}} = E_{\mathrm{pump}}-0.21$ meV. A spatial filter from $y=5$ $\mu$m to $y=30$ $\mu$m is used to isolate the light emission from the downstream region, where $y=0$ corresponds to the position of the defect. (c) Measured excitation spectrum in the downstream region obtained from the position of the $k$-space peak (b) with (red circles) and without (black squares) pump. The solid red line shows the Bogoliubov spectrum of excitations in the downstream region, calculated by using the positive branch of Eq.~S4 with the measured $k_{\rm fluid}$ and $E_{\rm int}$; the dashed red line shows the corresponding negative branch. The black dash-dotted line shows the single-particle spectrum within the parabolic approximation, to which the Bogoliubov dispersion reduces in the absence of the pump. (d) Bogoliubov dispersion in the upstream region, calculated by inserting the values of $k_{\rm fluid}$ and $E_{\rm int}$ measured at $p = 100$~mW into Eq.~S4.}}
\end{figure}

\section{Pump-probe measurements in the downstream region}

Polariton interactions in the black hole regime result in significant renormalizations of the polariton branches, following the Bogoliubov dispersion~\cite{RMPIacopo}. In order to evidence this effect and confirm our interpretation in terms of speeds of sound of the polariton fluid, we measure the spectrum of excitations of the downstream fluid in a pump-probe experiment. The pump power is fixed at $p=100\,mW$ in the upstream region in the conditions for the realization of the acoustic horizon.
A second cw Ti:Sapphire laser at a slightly different energy and weak power $p_{\mathrm{probe}}=5$ mW, playing the role of a probe, is focused onto a $30$ $\mu$m spot in the downstream region [see Fig.~\ref{S3}(a)]. Polariton emission at the energy of the probe laser is collected in transmission geometry. 

As suggested in a slightly different context~\cite{Wouters}, the probe laser is used here to excite a propagating wave on top of the polariton fluid, similar to pouring a cup of water on top of a flowing river.  Thus, the spectrum of excitations of the fluid can be reconstructed by measuring  the wave vector $k$ of the propagating wave excited by the probe and the wave vector $k'$ of the part of the wave reflected against the defect, at different energies of the probe laser, $E_{\mathrm{probe}}$. The emission of the background fluid is filtered out with the use of a spectrometer. Figure~\ref{S3}(b) depicts the probe signal as a function of the wave vector, with $E_{\mathrm{probe}} = E_{\mathrm{pump}}-0.21$ meV. We observe that the excited wave propagates at smaller wave vectors in the presence of the pump. Indeed, this effect is due to the renormalization of the spectrum of excitations due to the presence of the pump fluid in the nonlinear regime. The observation of excited modes below the pump energy is a clear evidence of the supersonic regime~\cite{Iacopofluid}. By repeating the same measurements at different $E_{\mathrm{probe}}$ values, we are able to reconstruct the dispersion in the downstream region, both in the linear (i.e. without pump) and nonlinear (i.e. in the presence of the pump) regimes [see Fig.~\ref{S3}(c)]. The dispersion in the linear regime corresponds well to the parabola $E(k_y)=E_0 + \hbar^2 k_y^2/2m$, while the dispersion in the nonlinear regime follows the expected positive branch of the Bogoliubov dispersion~\cite{RMPIacopo,Dario}:
\begin{widetext}
\begin{equation}
 E_{\pm}(k_y)= E_{\mathrm{pump}} \pm \frac{\hbar^2}{2m}\sqrt{(k_y-k_{\mathrm{fluid}})^2\left[(k_y-k_{\mathrm{fluid}})^2+\frac{2m}{\hbar}gn\right]} 
 +\frac{\hbar^2k_{\mathrm{fluid}}(k_y-k_{\mathrm{fluid}})}{m} \, , \label{Bogoliubov}
\end{equation}
\end{widetext} 
where $k_{\mathrm{fluid}}=0.52$ $\mu$m$^{-1}$ and $E_{\mathrm{int}}=\hbar g n=50$ $\mu$eV are the wavevector and interaction energy of the downstream fluid, respectively, corresponding to a pump power  $p=100$ mW [see Figs.~2(f) and 3(a) of the main text]. The same experiment, but measured in the upstream region, is technically more difficult since the probe signal is washed out by the dominant fluid emission even with the use of a spectrometer. Nevertheless, we can calculate the excitation spectrum of the upstream fluid using the Bogoliubov dispersion~\ref{Bogoliubov} with the parameters measured at $p=100$ mW ($k_{\mathrm{fluid}} = 0.21$ $\mu$m$^{-1}$ and $\hbar g n=0.35$ meV) as shown in Fig.~\ref{S3}(d). This dispersion, calculated with the measured parameters presented in the main text, corresponds to that of a subsonic fluid, indeed.

\section{Estimation of the Hawking temperature}
From the interaction energy above 100~mW [Fig.~3(a) of the main text], we estimate an average healing length of:
\begin{equation}
\xi \approx \frac{\hbar}{2\sqrt{m}}\left(\frac{1}{\sqrt{E_{\mathrm{int}}^{u}}}+\frac{1}{\sqrt{E_{\mathrm{int}}^{d}}}\right) \approx 1.5\,\, \mu\mathrm{m}.
\end{equation}
where $E_{\mathrm{int}}^{u(d)}$ is the interaction energy in the upstream (downstream) region. Since the horizon ($d\approx 8$~$\mu$m, see Fig.~1(g) in the main text) is smoother than the healing length ($d\gg\xi$), a Hawking temperature can be estimated in the hydrodynamic approximation~\cite{FinazziParentani}: 
\begin{equation}
T_H\approx\frac{\hbar}{k_Bd}\frac{(v_d^2-c_d^2)-(v_u^2-c_u^2)}{c_u+c_d}\approx 3\,K\;\;
\end{equation}
where $v_{u(d)}$ and $c_{u(d)}$ are the measured fluid velocity and speed of sound in the upstream (downstream) region for an excitation power of 100 mW [see Fig.~3(b,c) in the main text]. In the case of polaritons, the characteristic energy with which this temperature should be compared is not the lattice or polariton temperature. Actually, polaritons are very weakly coupled to phonons in the semiconductor matrix and, when excited resonantly, their energy distribution is {at most} determined by their lifetime \cite{WertzNatphys:2010}. {Therefore, $T_{H}$ should be compared to the polariton lifetime.} In the present sample, the polariton lifetime corresponds to a linewidth of $\sim 50$ $\mu$eV, i.e. much smaller than the thermal energy associated to the expected Hawking radiation ($\sim 250$ $\mu$eV for 3~K). A difficulty in the observation of Hawking radiation could come from the shot noise in the photo-detection process, but this issue can be overcome with a sufficiently long integration time.

\section{ Numerical simulations I: density and velocity profiles }

We have modeled this experiment by a driven-dissipative Gross-Pitaevskii equation, or generalized non-linear Schrodinger equation. In particular, considering that we work very close to the bottom of the lower polariton dispersion, we restrict our model to a single field describing the lower polariton. Of course, it is implicit that the polariton is a composite quasi-particle, and the radiation detected is only related to its photonic component~\cite{RMPIacopo}.

\begin{figure}[t]
\begin{center}
\includegraphics[width=0.5\textwidth ]{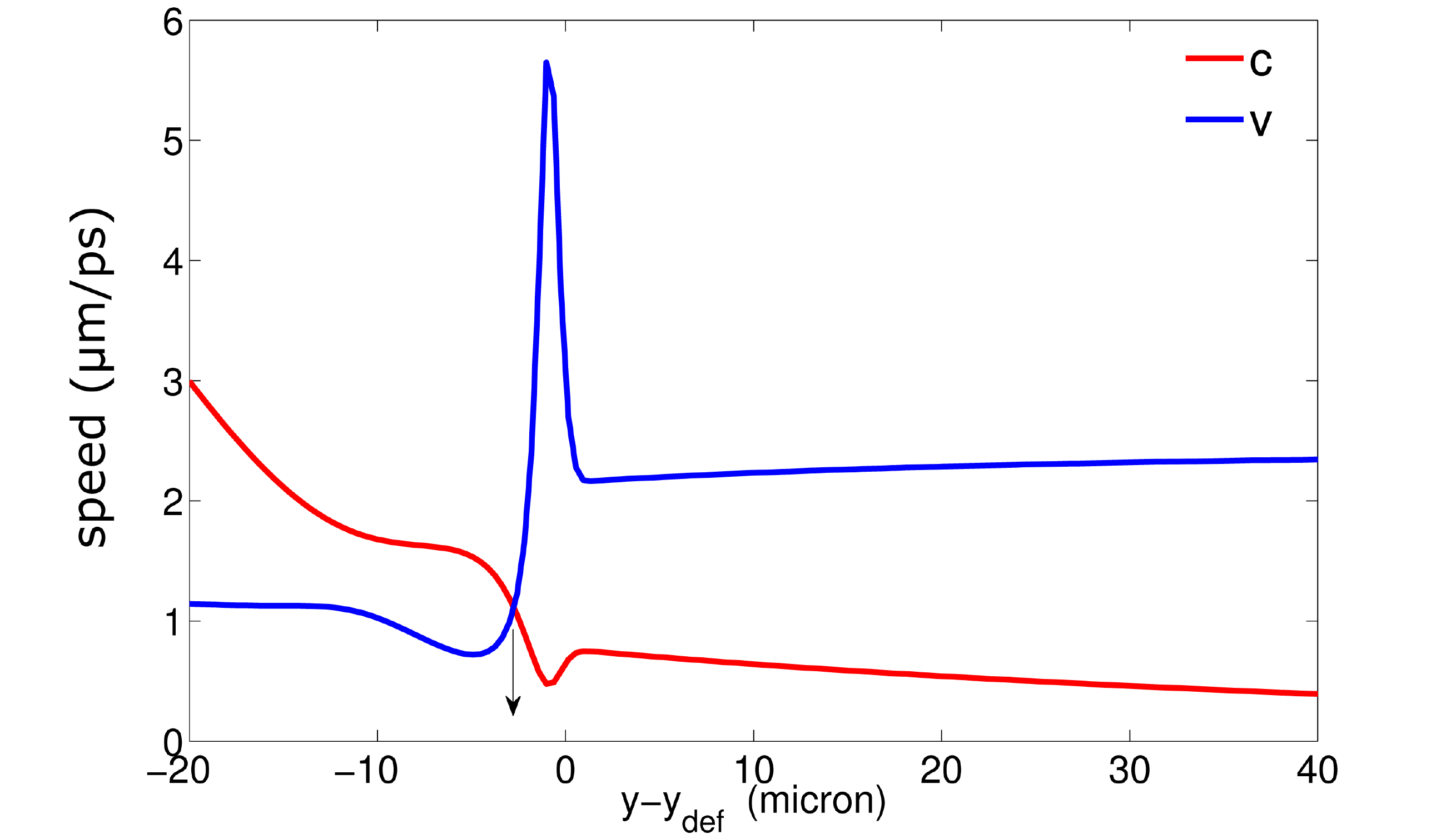}
\end{center}
\caption{\label{S4}{ Velocity profile as a function of propagation direction. The origin corresponds to the center position of the {microstructured} defect. The subsonic/supersonic boundary actually occurs about $2$ $\mu$m on its negative side (see arrow). }}
\end{figure}

In the dilute limit of weak polariton-polariton interactions, the dynamics of the system can be accurately captured by a mean-field treatment where the quantum polariton field, $\hat{\psi}$, is approximated by a classical field equal to its expectation value, $\phi=\langle \hat{\psi}\rangle$.
The evolution of such field is determined by a one-dimensional driven-dissipative nonlinear Schr\"odinger equation, or generalized Gross-Pitaevskii equation~\cite{RMPIacopo}:
\begin{widetext}
\begin{equation}\label{eq:mf_gpe}
{i\frac{d}{dt}  \phi(y,t) = \left[ \omega (-i \partial_y) + V(y) + g |\phi (y,t) |^2 - i \frac{\gamma }{2} \right] \phi (y,t) +F_{\mathrm{pump}}(y,t)  \, },
\end{equation}
\end{widetext}
which generalizes to the non-equilibrium context of polaritons the well-known Gross-Pitaevskii equation (GPE) of dilute Bose condensed gases~\cite{pitaevskii_stringari}.
In Eq.~\ref{eq:mf_gpe}, $V(y)$ is the spatially-dependent potential for polaritons propagating along the {microwire} axis, $y$ (i.e., describing the {microstructured} defect), $\omega (-i \partial_y)$ describes the polariton dispersion (approximated with a free-particle dispersion with constant effective mass, see main text), $g$ is the one-dimensional polariton-polariton interaction constant (assumed $0.3$ $\mu$eV$\cdot\mu$m in these simulations, as appropriate for a {microwire} width of $3$ $\mu$m), and {$F_{\mathrm{pump}}(y,t)=F_0(y)\,\exp[i(k_{\mathrm{pump}} y - \omega_{\mathrm{pump}} t)]$} describes the gaussian pump spot in continuous wave excitation regime.
 
The numerical solution of Eq.~\ref{eq:mf_gpe} allows to determine the fluid velocity and speed of sound along the wire axis, defined as $v(y)=\hbar \mathrm{Im}(\phi^* \partial_y \phi)/m$ and $c(y)=\sqrt{\hbar g n(y)/m}$, respectively. 
In particular, for the microstructure under investigation and at high pump power (corresponding to the $p=100$ mW experimental value, see the main text) the results are shown in Fig.~\ref{S4}, clearly evidencing the subsonic/supersonic transition of our analog black-hole event horizon.  

\section{ Numerical simulations II: spatial correlations }

The horizon created at the engineered defect of our {microwire} is particularly promising for the experimental detection of spontaneous Hawking radiation by vacuum field fluctuations at the  subsonic/supersonic boundary. An indirect experimental measure of such an emission can be performed by intensity correlation measurements at equal times. In order to estimate the feasibility of this experiment for the present horizon, we have calculated the density-density spatial correlations around the engineered defect, as shown in Fig.~4 of the main text.

To include the fluctuations of the polariton field around its mean value, we have employed a technique originally developed in a quantum optics context and then widely applied to dilute quantum fluids, the so-called {\em truncated Wigner approximation}~\cite{steel1998,sinatra}. This technique was recently extended to treat out-of-equilibrium quantum fluids of polaritons in~\cite{carusott2005prb,Dario}. In brief,  the dynamics of the quantum field problem can be described by a stochastic partial differential equation of the form:
\begin{widetext}
\begin{equation}
\label{eq:wigner}
i \, d\phi =\left[ \omega (-i \partial_y) + V(y) + g\left(|\phi |^2 - \frac{1}{\Delta y}\right)  - i \frac{\gamma}{2}  \right]  \phi \,dt + F_{\mathrm{pump}}(y,t) \, dt 
+  \frac{\sqrt{\gamma}}{\sqrt{4\,\Delta y}} \,d \xi  \, ,
\end{equation}
\end{widetext}
where $\Delta y$ is the spacing of the real-space grid, and $d \xi$ is a complex valued, zero-mean, independent Gaussian noise term with white noise correlation in both space and time, i.e. $\overline{d \xi^*(y,t)\,d \xi(y',t)}=2 \delta ({y}-{y}')\,dt$.
The equal-time spatial correlations of density fluctuations are calculated as
$g^{(2)}(y,y') = \langle \hat{\psi}^{\dagger} (y) \hat{\psi}^{\dagger} (y') \hat{\psi}(y') \hat{\psi}(y)  \rangle /  
\langle \hat{n}(y)  \rangle \langle  \hat{n} (y')  \rangle$, where $\langle \hat{n}(y)  \rangle \equiv \langle \hat{\psi}^{\dagger} (y) \hat{\psi}(y)  \rangle$, and the quantum expectation values can be calculated from Wigner averages over a large number of independent configurations ~\cite{carusott2005prb} obtained by sampling the stochastic evolution at different times spaced by $T_{\mathrm{s}} \gg 1/ \gamma$. In particular, the results shown in Fig.~4 have been obtained after averaging over $2 \times 10^5$ Montecarlo realizations.

\section{Spatial position of the maximum correlation signal}

The first calculations of the density correlation signal in~\cite{Balbinot,NJP} addressed the case of fluids whose density, flow velocity and interaction energy is spatially piecewise constant in both upstream and downstream regions. In such case, the intensity of the signal for the different processes was maximum on straight lines defined by same optical-path conditions. Locating the horizon at $y_{hor}=0$, for standard Hawking processes ($u-d_2$ in the notation of~\cite{PRA_gradino}), this condition reads
\begin{equation}
\frac{y}{-c^d+v^d}=\frac{y'}{-c^u+v^u}\;\;\;(y'<0<y)
\end{equation}
while it reads 
\begin{eqnarray}
\frac{y}{-c^d+v^d}=\frac{y'}{c^d+v^d}\;\;\;(0<y,y') \\
\frac{y}{c^d+v^d}=\frac{y'}{-c^u+v^u}\;\;\;(y'<0<y),
\end{eqnarray}
for other processes labelled as $d_1-d_2$ and $u-d_1$ in~\cite{PRA_gradino}, respectively.

When the flow velocity and/or the speed of sound are spatially varying, the above equations have to be modified into integrals. For instance, the standard Hawking processes are expected to give a maximum signal on the curve defined by the implicit integral equation
\begin{equation}
\int_{\approx 0}^y\!\frac{dY}{-c^d(Y)+v^d(Y)}= \int_{\approx 0}^{y'}\!\frac{dY}{-c^u(Y)+v^u(Y)}
\end{equation}
Of course, this equation is valid only within a geometrical optics approximation where the spatial variations are assumed to be slow as compared to the wavelength of the Hawking phonons. While this condition can be reasonably valid away from the horizon, exact determination of the lower integration limit in the vicinity of the horizon point at $y=0$ goes beyond this approximation. In Fig.~4 of the main text, the integration limits have been determined by hand, so to optimize agreement with the numerical result. In spite of this arbitrariness, the green line is in good agreement with the main fringe of standard Hawking correlations, i.e. the
$u-d_2$ feature of Ref.~\onlinecite{PRA_gradino}. A similar calculation has been performed with comparable success to obtain the black line for the $d_1-d_2$ feature of Ref~\onlinecite{PRA_gradino}. The appearance of additional fringes parallel to the main ones (barely visible in~\cite{PRA_gradino}) can be explained in terms of the strong curvature of the $k_y < 0$ part of the Bogoliubov dispersion in the downstream region, and from the complex density and velocity profiles in the horizon region~\cite{Zapata}.